\documentclass[11pt,a4paper]{article}
\usepackage[hyperref]{acl2021}
\usepackage{times}
\usepackage{latexsym}
\usepackage{graphicx}
\usepackage{fixltx2e}

\usepackage{microtype}

\aclfinalcopy 
\def\aclpaperid{3370} 

\title{Privacy at Scale: Introducing the PrivaSeer Corpus \\ of Web Privacy Policies}

\author{Mukund Srinath \qquad Shomir Wilson \qquad C. Lee Giles \\
College of Information Sciences and Technology \\
Pennsylvania State University \\
University Park, PA, USA \\
\texttt{\{mukund, shomir, clg20\}@psu.edu}}

\date{}

\begin{document}
\maketitle

\begin{abstract}
Organisations disclose their privacy practices by posting privacy policies on their websites. Even though internet users often care about their digital privacy, they usually do not read privacy policies, since understanding them requires a significant investment of time and effort. Natural language processing has been used to create experimental tools to interpret privacy policies, but there has been a lack of large privacy policy corpora to facilitate the creation of large-scale semi-supervised and unsupervised models to interpret and simplify privacy policies. Thus, we present the PrivaSeer Corpus of 1,005,380 English language website privacy policies collected from the web. The number of unique websites represented in PrivaSeer is about ten times larger than the next largest public collection of web privacy policies, and it surpasses the aggregate of unique websites represented in all other publicly available privacy policy corpora combined. We describe a corpus creation pipeline with stages that include a web crawler, language detection, document classification, duplicate and near-duplicate removal, and content extraction. We employ an unsupervised topic modelling approach to investigate the contents of policy documents in the corpus and discuss the distribution of topics in privacy policies at web scale. We further investigate the relationship between privacy policy domain PageRanks and text features of the privacy policies. Finally, we use the corpus to pretrain PrivBERT, a transformer-based privacy policy language model, and obtain state of the art results on the data practice classification and question answering tasks. 
\end{abstract}

\section{Introduction}

A privacy policy is a legal document that an organisation uses to disclose how they collect, analyze, share, and protect users' personal information. Legal jurisdictions around the world require organisations to make their privacy policies readily available to their users, and laws such as General Data Protection Regulation (GDPR) and California Consumer Privacy Act (CCPA) place specific expectations upon privacy policies. However, although many internet users have concerns about their privacy \cite{madden_privacy_2017}, most fail to understand privacy policies \cite{meiselwitz2013readability}. Studies show that privacy policies require a considerable investment in time to read \cite{obar2018biggest} and estimate that it would require approximately 200 hours to read all the privacy policies that an average person would come across every year \cite{mcdonald2008cost}.

Natural language processing (NLP) provides an opportunity to automate the extraction of salient details from privacy policies, thereby reducing human effort and enabling the creation of tools for internet users to understand and control their online privacy. Existing research has achieved some success using expert annotated corpora of a few hundred or a few thousand privacy policies \cite{wilson2016creation, zimmeck2019maps, ramanath2014unsupervised}, but issues of accuracy, scalability and generalization remain. More importantly, annotations in the privacy policy domain are expensive. Privacy policies are difficult to understand and many tasks such as privacy practice classification \citep{wilson2016creation}, privacy question answering \citep{ravichander2019question}, vague sentence detection \citep{lebanoff2018automatic}, and detection of compliance issues \citep{zimmeck2019maps} require skilled legal experts to annotate the dataset. In contrast, approaches involving large amounts of unlabeled privacy policies remain relatively unexplored. 

Modern robust language models, such as transformer-based architectures, benefit from increasingly large training sets. These models can be used on downstream tasks \citep{devlin2019bert}  to improve performance. Results have shown that in-domain fine tuning of such pre-trained language models have produced a significant boost in performance on many tasks \cite{gururangan2020don} in a variety of domains, suggesting a need for a larger collection of privacy policies to enable similar results in the privacy domain.

To satisfy the need for a much larger corpus of privacy policies, we introduce the PrivaSeer Corpus of 1,005,380 English language website privacy policies. The number of unique websites represented in PrivaSeer is about ten times larger than the next largest public collection of web privacy policies~\cite{amos2020privacy}, and it surpasses the aggregate of unique websites represented in all other publicly available web privacy policy corpora combined. We describe the corpus creation pipeline, with stages including a web crawler, language detection, document classification, duplicate and near-duplication removal, and content extraction. We then analyse the lengths and top level distribution of the privacy policies in the corpus and use topic modelling to explore the component topics. Subsequently, we pretrain PrivBERT, a transformer-based language model, using the corpus and evaluate it on data practice classification and question answering tasks. We release the corpus, a search engine for the corpus~\citep{srinath2021privaseer}, the document collection pipeline, and a language model to support further research in the privacy domain.\footnote{All artifacts are available at \url{https://privaseer.ist.psu.edu/}.} 

\section{Related Work}


Prior collections of privacy policy corpora have led to progress in privacy research. \citet{wilson2016creation} released the OPP-115 Corpus, a dataset of 115 privacy policies with manual annotations of 23k fine-grained data practices, and they created a baseline for classifying privacy policy text into one of ten categories. The corpus was used to train models to extract opt-out choices from privacy policies \citep{sathyendra2016automatic}, to automatically identify policies on websites and find compliance issues \citep{story2019natural}, and to classify privacy practices and answer privacy related non-factoid questions \citep{harkous2018polisis}. 

Other corpora similar to OPP-115 Corpus have enabled research on privacy practices. The PrivacyQA corpus contains 1,750 questions and expert-annotated answers for the privacy question answering task \citep{ravichander2019question}. Similarly, \citet{lebanoff2018automatic} constructed the first corpus of human-annotated vague words and sentences in privacy policies and studied automatic vagueness detection. \citet{sathyendra2017identifying} presented a dataset and developed a model to automatically identify and label opt-out choices offered in privacy policies. Similarly, \citet{zimmeck2019maps} released a set of over 400k URLs to Android app privacy policy pages collected by crawling the Google Play store. \citet{amos2020privacy} collected privacy policies from around 130,000 websites from over two decades and analysed the evolution of the online privacy landscape. Finally, \citet{zaeemlarge} collected a corpus of around 100k privacy policies using the domains from DMOZ, a website which maintained categories of websites on the internet.  

Prior work in privacy and human-computer interaction establishes the motivation for studying these documents. Although most internet users are concerned about privacy \citep{madden_privacy_2017}, \citet{rudolph2018users} reports that a significant number do not make the effort to read privacy notices because they perceive them to be too time-consuming or too complicated \citep{obar2018biggest}. Responding to the opaqueness of these document, \citet{schaub2015design} introduced methods to ease the design of privacy notices and their integration, and \citet{kelley2010standardizing} designed and tested a ``privacy nutrition label" approach to present privacy information visually. Suggestions to improve the presentation of privacy information, have not been adopted by many organisations. Apple has begun displaying privacy labels in its app stores having collected the information from App developers; however, concise privacy information for websites remains an open problem.

\section{Document Collection}

\begin{figure*}[h]
\centering
\includegraphics[scale=0.215]{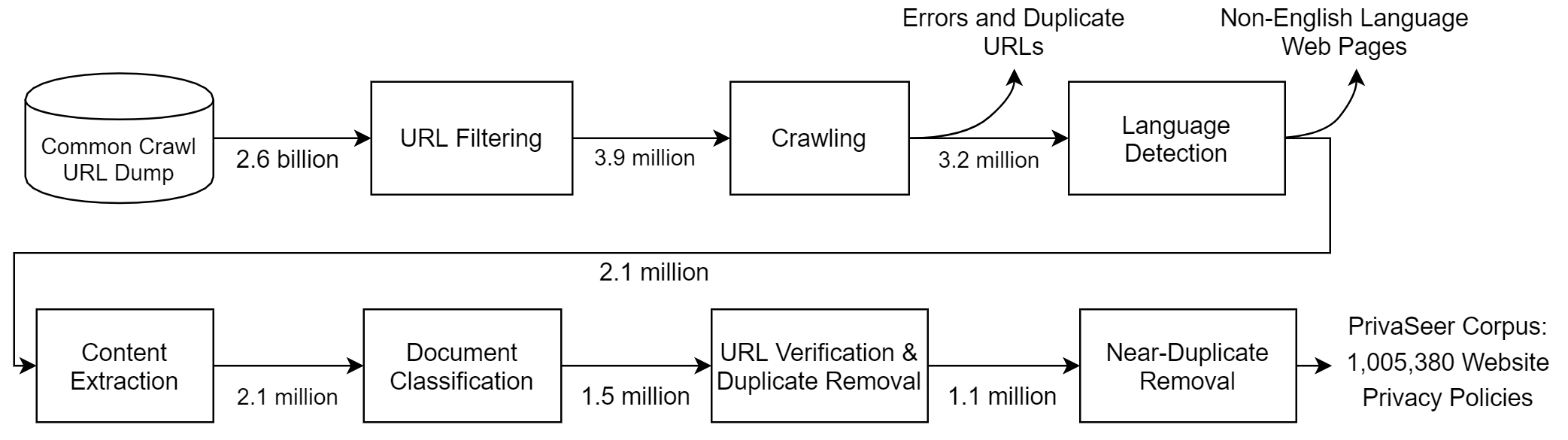}
\caption{Corpus creation pipeline}
\label{fig:pipeline}
\end{figure*}

To build the PrivaSeer corpus, we create a pipeline concentrating on focused crawling \citet{chakrabarti1999focused, diligenti2000focused} of privacy policy documents. We used Common Crawl,\footnote{\url{https://commoncrawl.org/}} described below, to gather seed URLs to privacy policies on the web. We filtered the Common Crawl URLs to gather a set of possible links to web site privacy policies. We then crawled the filtered set to obtain candidate privacy policy documents. The complete pipeline from the Common Crawl URL dump to the gold standard privacy policy corpus is in Figure \ref{fig:pipeline}. 


The Common Crawl Foundation has been releasing large monthly internet web crawls along with their web graphs since 2008. Monthly crawl archives provide a ``snapshot of the web" by including re-crawls of popular domains and crawls of new domains. 
We downloaded the URL dump of the May 2019 archive.\footnote{\url{https://commoncrawl.s3.amazonaws.com/crawl-data/CC-MAIN-2019-22/cc-index.paths.gz}} Common Crawl reports that the archive contains 2.65 billion web pages or 220 TB of uncompressed content which were crawled between 19th and 27th of May, 2019. We applied a selection criteria on the downloaded URL dump to filter the URLs of likely privacy policy pages.

Due to legal requirements, organizations typically include a link to their privacy policy in the footer of the website landing page commonly with the names \textit{Privacy Policy}, \textit{Privacy Notice}, and \textit{Data Protection}.
We selected those URLs which had the word ``privacy" or the words ``data" and ``protection" from the Common Crawl URL archive. We were able to extract 3.9 million URLs that fit this selection criterion. Informal experiments suggested that this selection of keywords was optimal for retrieving the most privacy policies with as few false positives as possible. To find the accuracy of this technique, we manually examined 115 English language website landing pages and their privacy policy URLs from the OPP-115 Corpus \citep{wilson2016creation} since it was built to cover the diverse distribution of privacy policies on the web, in terms of website popularity and sector of commerce. We found that out of 115 websites, 4 websites did not have their privacy policy links either on the landing page or one hop from the landing page and 5 other websites did not satisfy our URL selection criteria. Thus, our crawling technique would cover about 92.17\% $\pm$ 6.51\% of English privacy policies on the web with a 95\% confidence interval.


We crawled the 3.9 million selected URLs using Scrapy\footnote{\url{https://scrapy.org/}} for about 48 hours between the 4th and 10th of August 2019, for a few hours each day. 3.2 million URLs were successfully crawled, henceforth referred to as \textit{candidate} privacy policies, while 0.4 million led to error pages and 0.3 million URLs were discarded as duplicates. 


\section{Document Filtering}


\textbf{Language Detection.} We focused on privacy policies written in the English language, to enable comparisons with prior corpora of privacy policies. To identify the natural language of each candidate document, we used the open-source Python package Langid \citep{lui2012langid}. Langid is a Naive Bayes-based classifier pretrained on 97 different languages, designed to achieve consistently high accuracy over a wide range of languages, domains, and lengths of text. 
The complete set of documents was divided into 97 languages and an unknown language category. We found that the vast majority of documents were in English. We set aside candidate documents that were not identified as English by Langid and were left with 2.1 million candidates.



\textbf{Content Extraction.} Manual inspection of the English language web pages showed that they included content other than the main text: often they had a header, a footer, a navigation menu, and banners. We refer to this extra content in a web page as \textit{boilerplate}. Boilerplate draws away from the focus of the main content in a web page and therefore various techniques have been used to remove boilerplate from web pages \citep{gottron2007evaluating, weninger2016web}. After manual comparison of a number of content extraction tools, we used the open-source Python package \textit{boilerpipe} \citep{kohlschutter2010boilerplate} due to its superior performance. Boilerpipe effectively strips web pages of boilerplate using shallow text features, structural features and density based features. 


\textbf{Document Classification.} Some of the web pages in the English language candidate document set may not have been privacy policies and instead simply satisfied our URL selection criteria. To separate privacy policies from other web documents we used a supervised machine learning approach. Two researchers in the team labeled 1,600 randomly selected candidate documents based on a preset scheme in consultation with a privacy expert. While both the researchers had substantial prior experience with privacy policies, the privacy expert was consulted to eliminate uncertainty in the annotations of a few documents. Lack of agreement in the annotations occurred for six documents, which were settled by discussion with the expert. 
Out of 1,600 documents, 1,145 were privacy policies and 455 were not privacy policies. 

We trained four supervised machine learning models using the manually labelled documents with features extracted from the URLs and the words in the web page. We trained three random forest models and fine-tuned a transformer based pretrained language model, namely RoBERTa \citep{liu2019roberta}. The three random forest models were trained on three different sets of features: one using the features extracted from the URL, one using the features extracted from the document content, and a combined model using features from both.

For the URL model, the words in the URL path were extracted and the tf-idf of each term was recorded to create the features \citep{baykan2009purely}. As privacy policy URLs tend to be shorter and have fewer path segments than typical URLs, length and the number of path segments were added as features. Since the classes were unbalanced, we over-sampled from the minority class using the synthetic minority over-sampling technique (SMOTE) \citep{chawla2002smote}. Similarly, for the document model, we used tf-idf features after tokenizing the document using a regex tokenizer and removing stop words. The combined model was a combination of the URL and document features. 

To train the RoBERTa model on the privacy policy classification task, we used the sequence classification head of the pretrained language model from HuggingFace \citep{Wolf2019HuggingFacesTS}. We used the pretrained RoBERTa tokenizer to tokenize text extracted from the documents. Since Roberta accepts a maximum of 512 tokens as input, only the first 512 tokens of text from the documents were used for training while the rest was discarded. As shown in the analysis section, the average length of a privacy policy in terms of the number of words is 1,871. Thus 512 tokens would take into account about a fourth of an average privacy policy.

The 1,600 labelled documents were randomly divided into 960 documents for training, 240 documents for validation and 400 documents for testing. Using 5-fold cross-validation, we tuned the hyperparameters for the models separately with the validation set and then used the held-out test set to report the test results. Due to its size, it was possible for the held out test set to have a biased sample. Thus we repeated the sampling and training processes with a 5-fold cross-validation approach. Table \ref{classification-results} shows performance of the models after the results from test sets were averaged. Since the transformer based model had the best results, we ran it on all the the candidate privacy policies. Out of 2.1 million English candidate privacy polices, 1.54 million were classified as privacy policies and the rest were discarded. 



\begin{table}[h]
\centering
\begin{tabular}{|c|c|c|c|}
\hline \textbf{Model} & \textbf{Precision} & \textbf{Recall} & \textbf{F1} \\ \hline
URL Based & 0.88 & 0.89 & 0.88 \\ 
Document Based & 0.93 & 0.95 & 0.94 \\ 
Combined & 0.94 & 0.97 & 0.95 \\ 
RoBERTa & 0.97 & 0.98 & 0.97 \\
\hline
\end{tabular}
\caption{\label{classification-results} Document classification}
\end{table}

\textbf{URL Cross Verification.} Legal jurisdictions around the world require organisations to make their privacy policies readily available to their users. As a result, most organisations include a link to their privacy policy in the footer of their website landing page. In order to focus PrivaSeer Corpus on privacy policies that users are intended to read, we cross-verified the URLs of the privacy policies in our corpus with those that we obtained by crawling the homepages (landing page) of these domains. Between the 8th and 10th November 2019, we crawled the landing pages and pages one hop from the landing pages for all the domains of the URLs in our corpus. We then gathered the URLs satisfying our selection criteria and cross-verified them with the URLs in our existing corpus. After cross-verifying the URLs, we were left with a set of 1.1 million web pages.


\textbf{Duplicate and Near-Duplicate Detection.} Examination of the corpus revealed that it contained many duplicate and near-duplicate documents. We removed exact duplicates by hashing all the raw documents and discarding multiple copies of exact hashes. Through manual inspection, we found that a number of privacy policies from different domains had very similar wording, differing only by the organisation or website name. We reason that this similarity could be due to the use of privacy policy templates or generators. We also found abundant examples of near-duplicate privacy policies on the same website. We reason that this similarity could be due to the presence of archived versions of privacy policies on the website. Since we aimed to collect a comprehensive corpus of contemporary policies, we only removed similar policies (near-duplicates) from same domain domains.


To remove near-duplicates from within the same domain we used \textit{Simhashing} \citep{charikar2002similarity}. Simhashing is a hashing technique in which similar inputs produce similar hashes. After creating shingles \citep{broder1997syntactic} of size three, we created 64 bit document Simhashes and measured document similarity by calculating the \textit{Hamming distance} \citep{manku2007detecting} between document Simhashes of privacy policies within the same domain. We then obtained a list of all pairs of similar documents based on a distance threshold (measured based on the number of differing bits) that was determined after manual examination of a number of pairs of privacy policies. We then filtered the duplicates based on a greedy approach retaining policies that were longer in length. The remaining documents comprised the corpus. 




\section{Corpus Analysis}

The PrivaSeer Corpus consists of 1,005,380 privacy policies from 995,475 different web domains. Privacy policies in this corpus have a mean word length of about 1,871 words and range between a minimum of 143 words and a maximum of 16,980 words. 
The corpus contains policies from over 800 different top level domains (TLDs). \textit{.com}, \textit{.org}, and \textit{.net} make up a major share of the corpus covering 63\%, 5\% and 3\% respectively. Country-level domains like \textit{.uk}, \textit{.au}, \textit{.ca} and \textit{.du} show the geographic variety of the sources of the corpus covering 12\%, 4\%, and 2\% respectively. The distribution of popular TLDs (.com, .org, .net) roughly matches internet TLD trends suggesting that the corpus contains a random sample of internet web domains. Moreover, CommonCrawl release statistics estimating the representativeness of monthly crawls which support the claim that monthly crawl archives and in turn the PrivaSeer Corpus are a representative sample of the web. 
In addition to monthly crawl dumps, Common Crawl releases web graphs with PageRanks of the domains in a crawl. The PageRank values were calculated from the web graph using the Gauss-Seidel algorithm \citep{arasu2002pagerank}. PageRank values can be used as a substitute for popularity where higher values suggest more popular domains.

\textbf{Readability.} Readability of a text can be defined as \textit{the ease of understanding or comprehension due to the style of writing} \citep{klare1963measurement}. Along with length, readability plays a role in internet users' decisions to either read or ignore a privacy policy \citep{ermakova2015readability}. While prior studies on readability have shown that privacy policies are difficult to understand for the average internet user, they were conducted using small samples of policies and therefore may not be representative of the larger internet \citep{fabian2017large}. While there are a variety of readability metrics, we calculated the readability of the policies in the corpus using the Flesh-Kincaid Grade Level (FKG) metric for comparison with prior literature and since it is the the most widely used metric. The FKG metric presents the readability score as a U.S. grade level. We obtained a mean FKG score of 14.87 and a standard deviation of 4.8. This score can be interpreted as an average of 14.87 years of education in the U.S. (roughly two years of college education) is required to understand a privacy policy. In contrast, \citet{fabian2017large} found that the mean FKG score is 13.6 when they conducted an analysis of readability of privacy policies using 50k documents. 

\textbf{Topic Modelling.} Topic modelling is an unsupervised machine learning method that extracts the most probable distribution of words into topics through an iterative process \citep{wallach2006topic}. We used topic modelling to explore the distribution of themes of text in our corpus. Topic modelling using a large corpus such as PrivaSeer helps investigate the themes present in privacy policies at web scale and also enables the comparison of themes that occur in the rapidly evolving online privacy landscape. We used Latent Dirichlet Allocation (LDA), as our approach to topic modelling \citep{blei2003latent}. Since LDA works well when each input document deals with a single topic, we divided each privacy policy into its constituent paragraphs \citep{sarne2019unsupervised}, tokenized the paragraphs using a regex character matching tokenizer and \textit{lemmatized} the individual words using NLTK's WordNet lemmatizer. We experimented with topics sizes of 7, 8, 9, 10, 11, 13 and 15. We manually evaluated the topic clusters by inspecting the words that most represented the topics. We noted that the cohesiveness of the topics decreased as the number of topics increased. We chose a topic size of 9, since larger topic sizes produced markedly less coherent topics.

For each topic, we identified a corresponding entry from the OPP-115 annotation scheme \citep{wilson2016creation}, which was created by legal experts to label the contents of privacy policies. While \citet{wilson2016creation} followed a bottom-up approach and identified different categories from analysis of data practices in privacy policies, we followed a top-down approach and applied topic modelling to the corpus in order to extract common themes for paragraphs. The categories identified in the OPP-115 Corpus can be found in Table \ref{practice-classification}. 

We found that two LDA topics contained vocabulary corresponding with the OPP-115 category \textit{First Party Collection/Use}, one dealing with purpose and information type collected and the other dealing with collection method. Two LDA topics corresponded with the OPP-115 category \textit{Third Party Sharing and Collection}, one detailing the action of collection, and one explaining its purpose and effects(advertising and analytics). One of the LDA topics exclusively comprised of vocabulary related to cookies which could be related to both first party or third party data collection techniques. The OPP-115 categories \textit{Privacy Contact Information}, \textit{Data Security} and \textit{Policy Change} appeared as separate topics while a topic corresponding to the OPP-115 category \textit{International and Specific Audiences} appeared to be primarily related to European audiences and GDPR.  


\begin{figure}[h]
\centering
\includegraphics[scale=0.45]{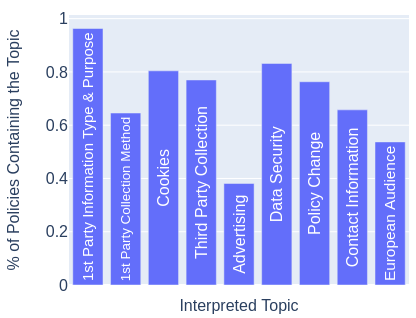}
\caption{Topic distribution}
\label{fig:topic-dist}
\end{figure}

It is likely that the divergence between OPP-115 categories and LDA topics comes from a difference in approaches: the OPP-115 categories represent themes that privacy experts expected to find in privacy policies, which diverge from the actual distribution of themes in this text genre. Figure \ref{fig:topic-dist} shows the percentage of privacy policies in the corpus that contain each topic. From the figure we see that information regarding the type and purpose of data collected by first and third party sources are the most common topics. About 77\% of policies contain language regarding third parties. This is consistent with prior research on third party data collection \citep{libert2018automated}. In contrast, language regarding advertising and analytics appears in only 38\% of policies in the corpus. Topics corresponding to data security, policy change and contact information also occur in a majority of privacy policies. Language corresponding to the GDPR and European audiences appears in 55\% of policies. A study of the distribution of privacy policy topics on the web is important since they inform us about real-world trends and the need for resource allocation to enforce of privacy regulations. 

\begin{figure}[h]
\centering
\includegraphics[scale=0.45]{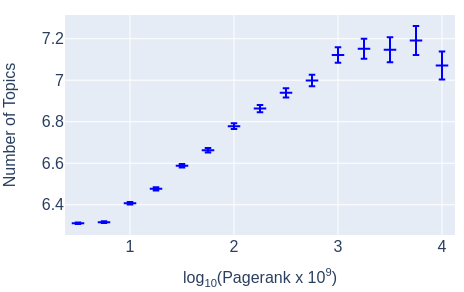}
\caption{Relationship between number of topics and privacy policy domain PageRank value}
\label{fig:topic-rank}
\end{figure}

Figure \ref{fig:topic-rank} shows how the number of topics in privacy policies vary with respect to the PageRank value. The whiskers in the plot represent the 95\% confidence interval of the means of the number of topics in the privacy policies in each PageRank value bin. The PageRank values were binned with a constant value of 0.25 such that each bin had at least 1k privacy policies. The plot suggests that more popular domains (as given by PageRank value) tend to address a greater number of topics in their privacy policies. This behaviour is consistent with manual inspection and is likely due to a larger and more diverse user base as well as the greater levels of regulatory scrutiny that accompany it in the case of more popular domains. For example, popular organisations tend to be multinational thereby requiring to address privacy laws from multiple jurisdictions such as GDPR from the European Union and CCPA from the United States. We found a similar pattern between privacy policy length and PageRank value thereby further supporting our claim that the more popular domain privacy policies tend to address a greater number of topics. In addition we found that readability and PageRank follow a similar pattern where privacy policies of more popular domains (as given by PageRank values) tend to be slightly more difficult to read. 

\section{PrivBERT}

In order to address the requirement of a language model for the privacy domain, we created PrivBERT. BERT is a contextualized word representation model that is  pretrained using bidirectional transformers \citep{devlin2019bert}. It was pretrained on the masked language modelling and the next sentence prediction tasks and has been shown to achieve state of the art results in many NLP tasks. RoBERTa improved upon the results achieved by BERT by making improvements to the training technique \citep{liu2019roberta}. We pretrain \textit{PrivBERT} starting with the pretrained RoBERTa\textsubscript{BASE} model (12 layers, 768 hidden size, 12 attention heads, 110M parameters). RoBERTa was trained on corpora of books, news articles, Wikipedia and social media comments and works well as a general purpose language model. Privacy policies written in legalese differ significantly in language when compared to the corpora used to train BERT and its variants, thereby prompting the need for a separate pretrained language model. Prior literature has shown that in-domain language models such as SciBERT \citep{beltagy2019scibert} and BioBERT \citep{lee2020biobert} perform significantly better on tasks in their respective domains. 

\begin{table*}[h]
\centering
\scalebox{0.95}{
\begin{tabular}{|c|c|c|c|c|c|c|c|c|}
\hline
\multicolumn{1}{|c|}{Label} & \multicolumn{1}{|c|}{Polisis} & \multicolumn{3}{|c|}{RoBERTa} & \multicolumn{3}{|c|}{PrivBERT} & \multicolumn{1}{|c|}{Support} \\ 
\hline
\multicolumn{1}{|c|}{} & \multicolumn{1}{|c|}{F1} & \multicolumn{1}{|c|}{P} & \multicolumn{1}{|c|}{R} & \multicolumn{1}{|c|}{F1} & \multicolumn{1}{|c|}{P} & \multicolumn{1}{|c|}{R} & \multicolumn{1}{|c|}{F1} & \multicolumn{1}{|c|}{} \\
\hline
First Party Collection and Use & 0.82 & 0.92 & 0.89 & 0.91 & 0.95 & 0.88 & \textbf{0.92} & 250 \\
Third Party Sharing and Collection & 0.82 & 0.86 & 0.91& 0.88 & 0.88 & 0.93 & \textbf{0.91} & 203 \\
User Choice/Control & 0.72 & 0.83 & 0.78 & 0.80 & 0.87 & 0.79 & \textbf{0.83} & 77 \\
Privacy Contact information & 0.84 & 0.84 & 0.76 & 0.79 & 0.82 & 0.80 & \textbf{0.81} & 42 \\
Introductory/Generic & 0.73 & 0.88& 0.67 & 0.76 & 0.79 & 0.76 & \textbf{0.77} & 78 \\
Practice Not Covered & 0.13 & 0.50 & 0.32 & 0.39 & 0.65 & 0.44 & \textbf{0.52} & 25 \\
Data Security & 0.75 & 0.91 & 0.75 & 0.82 & 0.94 & 0.80 & \textbf{0.86} & 40 \\
User Access, Edit and Deletion & 0.70 & 0.72 & 0.88 & 0.79 & 0.80 & 0.88 & \textbf{0.84} & 24 \\
Policy Change & 0.88 & 0.79 & 0.90 & 0.84 & 0.87 & 0.95 & \textbf{0.91} & 21 \\
Do Not Track & 1.0 & 1.0 & 1.0 & 1.0 & 1.0 & 1.0 & \textbf{1.0} & 3 \\
International and Specific Audiences & 0.82 & 0.83 & 0.78 & 0.81 & 0.89 &  0.84 & \textbf{0.86} & 56\\
Data Retention & 0.40 & 0.80 & 0.57 & 0.67 & 0.83& 0.71 & \textbf{0.77} & 14 \\
\hline
\hline
Macro Averages & 0.71 & 0.82 & 0.77 & 0.79 & 0.86 & 0.82 & \textbf{0.83} & 833 \\
Micro Averages & 0.78 & 0.86 & 0.82 & 0.84 & 0.88 & 0.85 & \textbf{0.87} & 833 \\
\hline
\end{tabular}}
\caption{\label{practice-classification} Test performance comparison of three models on the data practice classification task (P:Precision, R: Recall)}
\end{table*}

We use the byte pair encoding tokenization technique utilized in RoBERTa and retain its cased vocabulary. We did not create a new vocabulary since the two vocabularies are not significantly different and any out-of-vocabulary words can be represented and tuned for the privacy domain using the byte pair encoding vocabulary of RoBERTa. We preprocessed the privacy policy documents to create sequences of a maximum length of 512 tokens. Inputs significantly shorter than the maximum length occasionally occurred since we did not create sequences that crossed document boundaries. We trained PrivBERT using dynamic masked language modelling \citep{liu2019roberta} for 50k steps with a batch size of 512 using the gradient accumulation technique on two NVIDIA Titan RTX for 8 days with a peak learning rate of 8e-5. Other hyperparameters were set similar to RoBERTa. 

\textbf{Finetuning PrivBERT.} We evaluated the performance of PrivBERT on two tasks: (i) Data practice classification (ii) Answer sentences selection.

For the data practice classification task, we leveraged the OPP-115 Corpus introduced by \citet{wilson2016creation}. The OPP-115 Corpus contains manual annotations of 23K fine-grained data practices on 115 privacy policies annotated by legal experts. To the best of our knowledge, this is the most detailed and widely used dataset of annotated privacy policies in the research community. The OPP-115 Corpus contains paragraph-sized segments annotated according to one or more of the twelve coarse-grained categories of data practices. We fine-tuned PrivBERT on the OPP-115 Corpus to predict the coarse-grained categories of data practices. We divided the corpus in the ratio 3:1:1 for training, validation and testing respectively. Since each segment in the corpus could belong to more than one category and there are twelve categories in total, we treated the problem as a multi-class, multi-label classification problem. After manually tuning hyperparameters, we trained the model with a dropout of 0.15 and a learning rate of 2.5e-5.

Table \ref{practice-classification} shows the results for the data practice classification task comparing the performance between RoBERTa, PrivBERT and Polisis \citep{harkous2018polisis}, a CNN based classification model. We report reproduced results for Polisis since the original paper takes into account both the presence and absence of a label while calculating the score for each label \citep{nejad2020establishing}. Due to the unbalanced nature of the dataset, we report the macro-average and micro-average scores. PrivBERT achieves state of the art results improving not only on the macro-average F1 score of RoBERTa by about 4\% but also improving on the F1 score for every category in the task.  

For the question answering task, we leveraged the PrivacyQA corpus \citep{ravichander2019question}. PrivacyQA consists of 1,750 questions about the contents of privacy policies from 35 privacy documents. While crowdworkers were asked to come up with privacy related questions based on public information about an application from the Google Play Store, legal experts were recruited to identify relevant evidence within respective privacy policies that answered the question asked by the crowdworkers. The goal of the question answering task is to identify a set sentences in the privacy policy that has information relevant to the question. \citet{ravichander2019question} divided the corpus into 1,350 questions for training and validation and 400 questions for testing where each question in the test set is annotated by at least three experts. We fine-tuned PrivBERT on the training set as a binary classification task on each question-answer sentence pair to identify if the sentence is evidence for the question or not. We trained the model with a dropout of 0.2 and a learning rate of 3e-6 with the positive and negative classes weighted in the ratio 8:1 during training. We used sentence level F1 as the evaluation metric as described by \citet{ravichander2019question}, where precision and recall are calculated by measuring the overlap between the predicted sentences and gold standard sentences. 

\begin{table}[h]
\centering
\begin{tabular}{|c|c|c|c|}
\hline \textbf{Model} & \textbf{Precision} & \textbf{Recall} & \textbf{F1} \\ \hline
BERT & 0.442 & 0.348 & 0.39 \\ 
PrivBERT & \textbf{0.483} & \textbf{0.424} & \textbf{0.452} \\ 
\hline
\end{tabular}
\caption{\label{qa-results} Performance comparison on the answer sentence selection task}
\end{table}

Table \ref{qa-results} shows the results for the answer sentence selection task comparing the performance between BERT and PrivBERT. Results from BERT are as reported by \citet{ravichander2019question}. PrivBERT achieves state of the art results improving on the results of BERT by about 6\%. PrivBERT therefore has been shown to achieve state of the art results in two significantly disparate tasks in the privacy domain suggesting that it can be used to improve the performance on various real-world tasks and application in the privacy domain. 

\section{Conclusion}


We created the PrivaSeer Corpus which is the first large scale corpus of contemporary website privacy policies and consists of just over 1 million documents. We designed a novel pipeline to build the corpus, which included web crawling, language detection, document classification, duplicate removal, document cross verification, content extraction, and near duplicate removal.

Topic modelling showed the distribution of themes of privacy practices in policies, corresponding to the expectations of legal experts in some ways, but differing in others. The positive relationship between PageRank of a domain and the number of topics covered in its policy indicates that more popular domains have a slightly greater coverage of these topics. We hypothesize that this is because more popular domains tend to have a larger and more diverse user base prompting the privacy policies to address laws from various jurisdictions.  

Prior research on the readability based on small corpora of privacy policies had found that they were generally hard to understand for the average internet user. Our large scale analysis using the Flesch-Kincaid readability metric was consistent with prior findings. We found that on average about 14.87 years or roughly about two years of U.S. college education was required to understand a privacy policy. 

We pretrained PrivBERT a language model for the privacy domain based on RoBERTa. We evaluated PrivBERT on the data practice classification and the question answering tasks and achieved state of the art results. 

We believe that the PrivaSeer Corpus will help advance research techniques to automate the extraction of salient details from privacy policies. PrivBERT will help improve results on various tasks in the privacy domain and help build stable and reliable privacy preserving technology. This should benefit internet users, regulators, and researchers in many ways.

\section*{Acknowledgments}

This work was partly supported by a seed grant from the College of Information Sciences and Technology at the Pennsylvania State University. We also acknowledge Adam McMillen for technical support.


\bibliographystyle{acl_natbib}
\bibliography{references}


\end{document}